\documentclass{nature}

\usepackage{graphicx} 
\usepackage[margin=0.7in]{geometry}
\usepackage{amsmath}
\usepackage{hyperref} 
\usepackage{multirow}
\usepackage{newtxtext}         
\usepackage{chemformula} 
\usepackage{lineno}
\usepackage{siunitx}
\usepackage{caption}
\usepackage{float}
\linenumbers            
\usepackage{booktabs}

\hypersetup{colorlinks, 
	linkcolor={blue!75!black!80!yellow},
	citecolor={blue!75!black!80!yellow}, 
	urlcolor={blue!75!black!80!yellow}
	}
\usepackage[cmintegrals]{newtxmath}
\usepackage{lipsum}
\usepackage{lineno}
\nolinenumbers
\usepackage{soul,xcolor}
\setstcolor{red}
\newcommand{\BostonCollege}{Department of Physics, Boston College, Chestnut Hill, MA, USA}

\title{Altermagnetism produces pair emission and absorption from dark excitons and magnons in \ch{La2O3Mn2Se2}}

\author{Birender Singh\(^1\), Xian Xu\(^2\), Sabrina R. Hatt\(^3\), Suvodeep Paul\(^4\), Yu-Mi Wu\(^5\), Chao-Chun Wei\(^6\), Violet Williams\(^1\), Kyung-Mo Kim\(^1\), Yihao Zhang\(^1\), Mohamed Shehabeldin\(^1\), Cameron Grant\(^1\), April Li\(^1\), Michael Geiwitz\(^1\), Xiaoyin Li\(^6\), Garrett E. Granroth\(^7\), Feng Liu\(^6\), Qiong Ma\(^1\), Benedetta Flebus\(^1\), Judy J. Cha\(^5\), Vinod M. Menon\(^4\), Benjamin A. Frandsen\(^3\), Diana Y. Qiu\(^2\), Huiwen Ji\(^6\) and Kenneth S. Burch\(^{1*}\)}

\begin{document}

\maketitle
    \begin{affiliations}
    \item \BostonCollege
    \item Department of Materials Science, Yale University, New Haven, CT, USA
    \item Department of Physics and Astronomy, Brigham Young University, Provo, UT, USA
    \item Department of Physics, City College of New York, New York, NY, USA
    \item Department of Materials Science and Engineering, Cornell University, Ithaca, NY, USA
    \item Department of Materials Science and Engineering, University of Utah, Salt Lake City, UT, USA
    \item Neutron Scattering Division, Oak Ridge National Laboratory, Oak Ridge, TN, USA

\noindent$^{*}$To whom correspondence should be addressed; E-mail: \href{mailto:ks.burch@bc.edu}{ks.burch@bc.edu}

\end{affiliations}
\begin{abstract}
Altermagnets' (AMs) non-relativistic spin splitting enables novel states, spintronic and magneto-optical devices, though their optical signatures remain elusive. Here, we report exciton-magnon emission and absorption: optical sidebands from a spin-forbidden dark exciton, a direct consequence of altermagnetic symmetry. Combined optical spectroscopy and first-principles calculations reveal that \ch{La2O3Mn2Se2} is an altermagnetic insulator, hosting a strongly bound, spin-forbidden dark exciton and a higher-energy bright exciton. Photoluminescence (PL) and absorption reveal mirror-image sidebands, Stokes-shifted in emission and anti-Stokes-shifted in absorption, symmetric about the dark exciton, whose energy shifts and spectral shapes match the magnon energy scale and density of states measured independently by inelastic neutron scattering. The PL intensity tracks the full equal-time spin-spin correlator, combining static and dynamical contributions, and rules out alternative processes. This directly couples PL to magnetism, with potential for magneto-optical devices. These results establish exciton-magnon spectroscopy as a new route for optically identifying and exploiting AMs.
\end{abstract}


Unlike conventional antiferromagnets (AFMs), altermagnets possess compensated magnetic moments on inequivalent lattice sites that are related neither by inversion (\(P\)) nor by translation. Instead, the magnetic sublattices are connected by a combination of time reversal (\(T\)) and typically a rotation (for example, \(C_{4z}\) or \(C_{3z}\)). This produces non-relativistic spin splitting of the bands enforced by crystal symmetry\cite{vsmejkal2022beyond, vsmejkal2022emerging, song2025}, sometimes referred to as crystalline-spin-valley-locking (CSVL)\cite{Ma.MultiAFM.2021,vila2025}. Such systems potentially combine the functionality of ferromagnets with antiferromagnetic order, offering new routes to physical phenomena useful in a variety of applications, such as piezomagnetism\cite{yershov2024,aoyama2024}, spintronics\cite{bai2024}, anomalous Hall effects\cite{reichlova2024,attias2024} and magneto-optical responses\cite{mazin2023,gray2024,sun2025,Wang2025}. The latter are particularly powerful; indeed, optics offers a table-top probe of magnetism and often imposes symmetry constraints that can separate specific contributions and types of magnetic order\cite{Kimel.OpticAM.2024}. Yet, studies of altermagnetism have largely focused on metallic systems\cite{krempasky2024,hariki2024x,lee2024,amin2024,wu2024valley,jiang2025,sun.KVS.2025,yang2025,zhou2025,wang2025Valley,zhang2024,reichlova2024,rial2024}, where altermagnetic order can be accessed through spin- and angle-resolved photoemission spectroscopy (ARPES) and transport measurements. 
Such approaches are unavailable in insulators, where identifying altermagnetic order has instead relied on techniques including X-ray magnetic circular dichroism (XMCD)\cite{hariki2024x}, neutron diffraction\cite{liu2024chiral,morano2025,sears2026}, and second-harmonic generation (SHG)\cite{ma2025}. Despite theoretical calculations suggesting numerous altermagnetic insulators\cite{gao2025Review,sodequist2024two,wang2026pent,wang2025real}, these approaches typically require special setups and remain challenging to implement in device geometries and nanostructures. A table-top optical probe would therefore be invaluable for identifying and classifying altermagnetic order in such systems, and one that couples directly to the underlying spin correlations could further provide a novel route toward magneto-optical devices.

In this vein, an underappreciated consequence of altermagnetism is that CSVL can produce an optically dark exciton state. As illustrated in Fig.~\ref{fig:Dark_exciton}a, unlike conventional AFMs with spin-degenerate bands, the non-relativistic spin splitting in AMs can result in the lowest-energy band-to-band transition, and hence the corresponding exciton, being spin-forbidden (\(\Delta S \neq 0\)). This mechanism is distinct from dark excitons arising from relativistic spin splitting due to spin-orbit coupling (SOC), which does not rely on magnetic order. Although optically forbidden, these excitonic states can become allowed through simultaneous magnon creation or annihilation. Such exciton-magnon transitions were first predicted in 1965\cite{tanabe1965} and subsequently observed in \ch{MnF2}\cite{greene1965,sell1968review}, now an established AM\cite{yuan2020,biniskos2025}.
In this process, the electric-dipole and exchange interactions act in concert to flip a spin, thereby enabling the otherwise forbidden optical transition. If the magnetic sublattices are related by \(PT\) symmetry, their degeneracy leads to a perfect cancellation of the corresponding quantum trajectories, enforcing zero exciton-magnon optical transitions\cite{tanabe1965}. AMs, however, explicitly break this \(PT\) symmetry, allowing non-zero exciton-magnon optical transitions. Thus, these transitions can provide an optical signature for their identification, while also enabling a direct route to strongly correlated spin and valley physics. This strongly correlated optical phenomenon, requiring both strong exciton-magnon coupling and the symmetry breaking unique to collinear AMs, has rarely been observed, and its key predictions remain untested. Specifically, a direct correlation of the optical transition energy with the magnon dispersion, and a correlation of the exciton-magnon emission intensity with the local spin-spin correlations.

The rare observation of exciton-magnon optical transitions results from the requirement of a judicious choice of material. Namely, one must combine strong magnetic excitations with an insulating ground state that enables optical access to these transitions. Thus, to demonstrate the utility of the exciton-magnon process as an optical signature of the altermagnetic state, we focus on \ch{La2O3Mn2Se2}, an insulator with weak spin-orbit coupling and a high-spin configuration conducive to a dark exciton. This material exhibits collinear G-type antiferromagnetic order (Fig.~\ref{fig:Dark_exciton}b), \(T_N\sim\)166~K\cite{PRB82}, and was recently suggested by some of us to host a Mott-insulating altermagnetic state on a Lieb lattice\cite{PRM2025Huiwen}. Here, we first establish the electronic structure of \ch{La2O3Mn2Se2} by combining optical spectroscopy and first-principles calculations (Fig.~\ref{fig:Dark_exciton}c). The photoluminescence (PL), absorption (\(\alpha\)), and photoluminescence excitation (PLE) measurements identify a spin-forbidden dark exciton at 1.56~eV. Both the PL (1.46~eV) and \(\alpha\) (1.66~eV) sidebands are Stokes- and anti-Stokes-shifted nearly symmetrically from the reference dark exciton energy. The overall spectra and observation of the spin-forbidden dark exciton are also consistent with first-principles calculations that further confirm the dark exciton originates from the CSVL. 

To determine if the large energy scale and spectral shapes result from magnons, we measured the magnon dispersion and fitted it with a model Hamiltonian. The resulting one- and three-magnon DOS convolved with the exciton bandwidth closely matches the energy shift and shape of the sidebands. The temperature evolution of the 1.46~eV PL reinforces the magnetic origin, as it quantitatively tracks the underlying spin-spin correlations obtained from the full equal-time spin-spin correlator. This behavior is distinct from typical PL and, together with the observed magnetic-field independence of PL up to 7~T, further suggests its exciton-magnon sideband origin. The nearly linear excitation power dependence and absence of saturation rule out defect-related emission, consistent with the high crystalline quality revealed by high-angle annular dark-field (HAADF) and integrated differential phase contrast (iDPC) scanning transmission electron microscopy (STEM). These observations also exclude alternative Franck-Condon and defect-related mechanisms\cite{ballhausen1962}. The observation of a spin-forbidden dark exciton and its strong electric-dipole transitions mediated by magnons provides direct optical evidence of altermagnetic order in \ch{La2O3Mn2Se2}. This exchange-mediated coupling further suggests a new composite state of spin-forbidden excitons and magnons unique to AMs. Our results establish exciton-magnon spectroscopy as a promising route to harnessing the spin-and-valley-locked states intrinsic to insulating AMs, extending valleytronic concepts established in transition-metal dichalcogenides to strongly correlated insulators. 

\section*{Electronic structure of \ch{La2O3Mn2Se2}}

\begin{figure}[]
\centering
\includegraphics[width=1\textwidth]{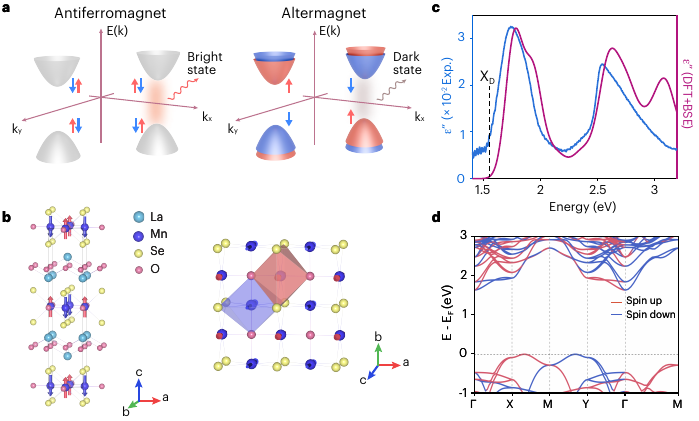}
\caption{\textbf{Dark exciton from altermagnetic spin splitting.} \textbf{a,} Spin-allowed optical transition in degenerate antiferromagnets (AFMs) (left) becomes spin-forbidden in altermagnets (AMs) (right) due to non-relativistic spin splitting of the bands. The red (blue) arrow indicates the up (down) spins. \textbf{b,} Crystal and magnetic structure of \ch{La2O3Mn2Se2}. Right: the \ch{Mn2OSe2} layer showing the two-sublattice sites via the red and blue octahedra. \textbf{c,} Imaginary part of the permittivity, \(\varepsilon''\) (blue), obtained from transmission and reflection, and theoretically calculated (pink). The vertical dashed line indicates the dark (\(X_D\)) exciton. \textbf{d,} Calculated spin-resolved band structure of \ch{La2O3Mn2Se2}, with red (blue) the spin up (down) projection along $\hat{z}$.}
\label{fig:Dark_exciton}
\end{figure}

Structurally, \ch{La2O3Mn2Se2} consists of a square lattice of two inequivalent Mn\(^{2+}\) ions in the \ch{MnSe4O2} octahedra\cite{PRB82,PRM2025Huiwen}, Fig.~\ref{fig:Dark_exciton}b. The Mn\(^{2+}\) ions have a high-spin (3\(d^5\)) valence configuration, and the two sublattices are connected by the combined real-space \(C_4\)-rotation and time-reversal (or \(C_2\) in spin-space symmetries), enforcing non-relativistic spin splitting and suggesting a spin-forbidden optical band gap. Previous density functional theory with Hubbard U (DFT+U) calculations find that the conduction band's (CB's) minimum and valence band's (VB's) maximum have opposite spin; however, these were not directly correlated with experimental measurements, so the value of Hubbard U was treated as a parameter leading to different ground states\cite{xu2012,PRM2025Huiwen,Roser2025}. The relatively high magnetic ordering temperature of \ch{La2O3Mn2Se2} (\(T_N \sim166\)~K), and the short-range magnetic correlations persisting up to room temperature\cite{PRM2025Huiwen} suggest large exchange interactions (\(\hat{V}\)), a key ingredient for sizeable spin splitting and strong exciton-magnon optical transitions. These characteristics uniquely identify \ch{La2O3Mn2Se2} as a candidate for exploring the optical signatures of AMs.

With this in mind, we first probed the optical transitions in \ch{La2O3Mn2Se2}. Figure~\ref{fig:Dark_exciton}c displays the absorptive, imaginary part of the dielectric function, (\(\varepsilon''\)), at 294 K, obtained by simultaneously fitting transmission and reflection spectra (Supplementary Fig.~1a) using the Kramers-Kronig-constrained variational dielectric function (vDF) method\cite{kuzmenko2005}. The measured data and fitting results for other optical constants are discussed in Supplementary Section~1. Two prominent features are observed in the \(\varepsilon''\) spectra at 1.74~eV and 2.54~eV. The high-energy, 2.54~eV band exhibits a broad asymmetric line shape and a low-energy tail around 2.34~eV, suggesting a continuum of optically allowed interband or excitonic transitions.

To investigate the origin of the absorption features in \(\varepsilon''\), we calculated the excitonic states of \ch{La2O3Mn2Se2} by solving the Bethe-Salpeter equation (BSE) within many-body perturbation theory\cite{deslippe2012,rohlfing2000}, starting from a DFT+U electronic structure\cite{anisimov1997}. Specifically, varying U strongly modifies the Mn \(d\)-states, and consequently the low-energy electronic structure\cite{xu2012,Roser2025}. Thus, the Hubbard U parameter was tuned to reproduce the experimentally observed 1.74~eV absorption feature in the BSE-calculated optical spectrum. The corresponding GW band structures for different values of U are compared in Supplementary Section~2. SOC was not included, as previous work has shown that it weakly affects the band structure and spin splitting\cite{PRM2025Huiwen}. The BSE-calculated optical spectrum (Fig.~\ref{fig:Dark_exciton}c, pink) captures the salient features of the experimental \(\varepsilon''\) spectra, albeit with a small blue shift. Specifically, the calculated spectrum exhibits broad bright excitonic peaks at 1.77~eV and 2.62~eV, the latter showing an asymmetric low-energy tail consistent with experiment. An additional feature predicted at 3.08~eV is absent experimentally, possibly owing to lifetime effects not included in the calculations.

To understand the origin of the peaks in the absorption spectrum, we examine the calculated electronic band structure of \ch{La2O3Mn2Se2}. This is shown in Figure~\ref{fig:Dark_exciton}d, with the spin-up (spin-down) projection along \(\hat{z}\) colored in red (blue). The band structure reveals the pronounced non-relativistic \(d\)-wave type momentum-dependent spin splitting with zero net magnetization, consistent with the altermagnetic ground state of \ch{La2O3Mn2Se2}. For U=2.2~eV, an indirect band gap of 1.63~eV is obtained, with the VB maximum located along the \(M-X\) (\(M-Y\)) high-symmetry directions and the CB minimum at \(\Gamma\). Additionally, calculations of the orbital-resolved band structure for different atoms (Supplementary Fig.~5) demonstrate that the lowest four CBs are mainly composed of Mn \(d_{z^2}\) and Mn \(d_{xy}\) orbitals, whereas the highest two VBs have contributions from Se \(p_z\), O \(p_{x,y}\), and Mn \(d_{x^2-y^2}\) orbitals near \(X/Y\) high-symmetry points, and from Se \(p_{x,y}\), O \(p_{x,y}\), and Mn \(d_{zx}/d_{zy}\) orbitals near the \(\Gamma\) point. 
Interestingly, the \(X~(Y)\) high-symmetry points with opposite spin polarization produce the lowest direct gap of 2.02~eV, thus suggesting a spin-forbidden transition between the highest VB and the lowest CB at the \(X/Y\)-points.

Next, we turn to identify the character of the excitonic states in the calculated absorption spectrum (Supplementary Section~2, Supplementary Figs.~6 and 7). The lowest-energy excitonic state is located at 1.39~eV and is an optically spin-forbidden dark exciton. This dark exciton is primarily composed of electrons from the two lowest CBs and holes from the two highest VBs in the vicinity of the \(X\) and \(Y\) high-symmetry points. These bands are relatively flat, resulting in a large exciton binding energy of \(\sim\) 0.63~eV with respect to the direct band gap. The peak in absorption at 1.77~eV corresponds to a bright excitonic state arising from the spin-allowed transitions between the third and fourth lowest CBs and the two highest VBs at \(X~(Y)\) with a binding energy of 0.39~eV with respect to the spin-allowed band-to-band transition. The large splitting of 0.38~eV between the lowest-energy spin-forbidden dark exciton and the bright exciton originates from the non-relativistic spin splitting of the CBs and their different exciton binding energies. Moreover, the broad peak at 2.62~eV originates from a distinct high-energy excitonic series arising from interband transitions near the \(\Gamma\)-point (Supplementary Fig.~7c). These are resonant states lying above the direct band gap at \(X\) and \(Y\), and a small amount of hybridization with the continuum of lower-energy states gives rise to the asymmetric low-energy tail.
\section*{Magnon-mediated optical transitions}

\begin{figure}[]
\centering
\includegraphics[width=0.95\textwidth]{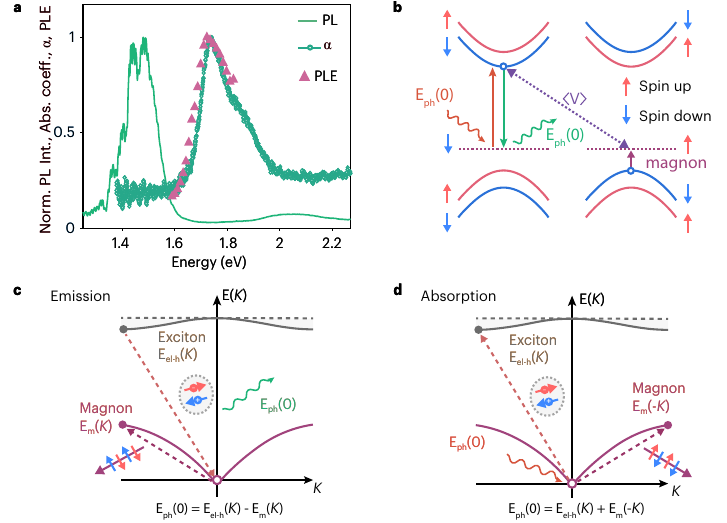}
\caption{\textbf{Optical excitations of \ch{La2O3Mn2Se2}.} \textbf{a,} The normalized PL (green), absorption coefficient, \(\alpha\) (green circles), and photoluminescence excitation (PLE) spectrum (pink triangles). \textbf{b,} Schematic illustration of exchange-assisted optical transitions involving the spin-forbidden dark exciton. In the absorption process (left), the incident photon simultaneously creates a dark exciton and a magnon through exchange coupling, \(\langle V\rangle\). In the emission process (right), radiative recombination of the dark exciton is accompanied by magnon creation, giving rise to the magnon-assisted Stokes-shifted emission. \textbf{c,d,} Momentum-space schematic of the proposed mechanism: photon emission occurs with simultaneous annihilation of the dark exciton and creation of a magnon, \(E_{\mathrm{ph}}(0)=E_{\mathrm{el-h}}(K)-E_{\mathrm{m}}(K)\) (c), optical absorption creates a dark exciton together with a magnon, \(E_{\mathrm{ph}}(0)=E_{\mathrm{el-h}}(K)+E_{\mathrm{m}}(-K)\) (d), where \(K\) is the exciton center-of-mass momentum. The dark exciton is located near the \(X/Y\) valleys (Fig.~1d and Supplementary Fig.~7) and is optically activated through exchange-assisted exciton-magnon coupling.}
\label{fig:Optical}
\end{figure}

To identify the experimental signatures of the spin-forbidden dark exciton, we directly probed the PL and absorption spectra of \ch{La2O3Mn2Se2}. Figure~\ref{fig:Optical}a displays the 14~K PL spectra (green), normalized to its maxima, together with \(\alpha\) (green circle). The spectra show a strong PL band centered at 1.46~eV, and a weak feature at 2.05~eV. Crucially, the 1.46~eV PL band is substantially shifted to lower energy when compared with the 1.74~eV bright exciton absorption feature. This behavior is confirmed across multiple samples, excitation lasers, and optical setups in three different labs (Supplementary Fig.~4a). Of particular note, PL measurements in samples of different thicknesses (Supplementary Fig.~4a) and \(k\)-space mapping (Supplementary Fig.~4b) identify the fast oscillations superimposed on the 1.46~eV PL band as Fabry-Pérot interference. The large Stokes shift of the 1.46~eV PL band is further verified in the PLE measurements. The PLE response was obtained at 14~K by monitoring the integrated intensity of the 1.46~eV PL band across an excitation range of 1.59~eV to 1.82~eV. The PLE spectrum (pink triangles, Fig.~\ref{fig:Optical}a) indirectly probes absorption via emission, independent of other absorption effects\cite{wang1993}. It peaks at 1.74~eV, well above the PL resonance (1.46~eV), consistent with excitation into higher-energy states followed by relaxation and emission from the lowest-energy spin-forbidden dark exciton. Beyond this, the PLE spectra closely follow the bright exciton absorption feature at 1.74~eV (Fig.~\ref{fig:Optical}a).

The large Stokes shift of the PL in \ch{La2O3Mn2Se2} could originate from a spin-forbidden excitonic transition enabled by an exciton-magnon process\cite{tanabe1965,sell1968review,eremenko1986}, from on-site \(d\)-\(d\) transitions through the Franck-Condon (F-C) mechanism\cite{ballhausen1962}, or from defect states. In the exciton-magnon process, as illustrated in Fig.~\ref{fig:Optical}b, an effective electric dipole, \(\hat{P}_{\mathrm{eff}}\), enables the otherwise spin-forbidden transition by coupling an electric-dipole-induced orbital excitation, \(\hat{E}\), to a change in the relative spin configuration through the exchange interaction\cite{tanabe1965,sell1968review}, \(\hat{V}\). This effect is governed by: \(\langle|\hat{P}_\mathrm{eff}|\rangle=\sum \langle|E|\rangle\langle |V|\rangle\), where the sum runs over the relevant intermediate orbital and spin states. If the two magnetic sites were connected by a center of inversion (i.e., \(PT\) symmetry is preserved), the degeneracy would produce an exact cancellation (\(\langle|\hat{P}_\mathrm{eff}|\rangle= 0\)), thereby preventing the exciton-magnon optical transitions\cite{tanabe1965}. A non-zero \(\langle\hat{P}_{\mathrm{eff}}\rangle\) therefore requires non-relativistic spin splitting with the symmetry that defines AMs. 

This symmetry-enabled, exchange-mediated coupling produces both exciton-magnon emission and absorption sidebands\cite{eremenko1986}, as schematically illustrated in Fig.~\ref{fig:Optical}c,d. Exciton-magnon emission proceeds via simultaneous annihilation of an exciton and creation of a magnon, producing a Stokes-shifted photon, \(E_{\mathrm{ph}}(0)=E_{\mathrm{el-h}}(K)-E_\mathrm{m}(K)\) (Fig.~\ref{fig:Optical}c), where \(K\) is the exciton center-of-mass momentum, and \(E_{\mathrm{ph}}(0)\), \(E_{\mathrm{el-h}}(K)\), and \(E_\mathrm{m}(K)\) denote the photon, spin-forbidden dark exciton, and magnon energies. Conversely, exciton-magnon absorption simultaneously creates an exciton and a magnon, giving an anti-Stokes-shifted transition, \(E_{\mathrm{ph}}(0)=E_{\mathrm{el-h}}(K)+E_\mathrm{m}(-K)\) (Fig.~\ref{fig:Optical}d)\cite{tanabe1965,sell1968review,eremenko1986}.
By contrast, the conventional F-C optical transitions involve vertical excitation to higher-energy manifolds that subsequently relax to the lowest-energy emissive state through electron–phonon coupling, enabling parity or spin mixing (via SOC). As such, the F-C Stokes shift in the PL is unrelated to the magnon energy; only the polarization is affected by magnetic correlations, and the total intensity is quenched by non-radiative pathways\cite{adachi2023I,Xiaodong2018}.

\begin{figure}[]
\centering
\includegraphics[width=1\textwidth]{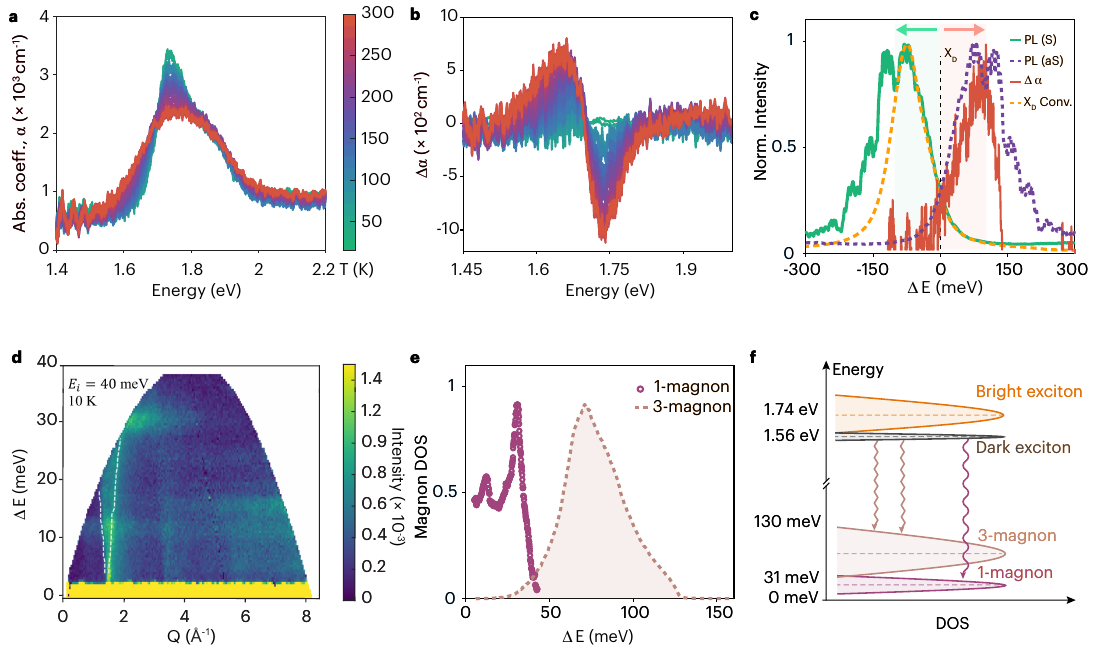}
\vspace{-4ex}
\caption{\textbf{Exciton-magnon optical transitions.} \textbf{a,} Temperature dependence of the absorption coefficient, \(\alpha\), from 15~K to 300~K. \textbf{b,} Differential absorption, \(\Delta\alpha=\alpha(T)-\alpha(15~\mathrm{K})\), highlighting the temperature dependent spectral weight redistribution. \textbf{c,} Stokes (S, solid green) and mirrored anti-Stokes (aS, dashed purple) 1.46~eV PL at 14~K, and differential absorption, \(\Delta\alpha\) at 300~K (red), plotted relative to dark exciton energy (\(X_D \sim\) 1.56 eV). The dashed orange curve represents the convolution of the measured one-magnon and three-magnon density of states (DOS) with an excitonic bandwidth of 57~meV. The calculated spectra quantitatively reproduce the experimentally observed optical sidebands. The vertical dashed line denotes the dark exciton energy, shifted to zero. \textbf{d,} Scattered neutron intensity \(S(Q,E)\) at 10~K using an incident neutron energy of 40~meV. The magnon excitations are most prominent between 1 and 2.5~\AA$^{-1}$. The dashed white lines represent the onset of the magnon plume (left line) and the maximum intensity within the magnon plume (right line). \textbf{e,} The measured one-magnon (pink circles) and calculated three-magnon (dashed line) density of states (DOS). \textbf{f,} Energy-level diagram of the exciton-magnon optical emission.} 
\label{fig:exciton-magnon}
\end{figure}

Hence, if the PL arises from exciton-magnon emission, the same dark exciton should also produce an anti-Stokes-shifted exciton-magnon absorption sideband below the bright exciton (Fig.~\ref{fig:Optical}d). Specifically, the two sidebands should mirror one another as they have equal and opposite shifts from a reference dark exciton energy\cite{sell1968review,eremenko1986}. However, a magnon-assisted absorption sideband is expected to emerge and gain spectral weight from the bright exciton upon heating, as the thermal population of magnon states enhances the joint exciton-magnon density of states\cite{eremenko1986}. Therefore, to isolate the magnon-assisted contribution from the dominant 1.74~eV bright exciton, we measured the temperature dependence of \(\alpha\) from 15 to 300~K. As shown in Fig.~\ref{fig:exciton-magnon}a, a transfer of spectral weight from the 1.74~eV bright exciton to lower energy is observed upon heating, while its high-energy edge remains nearly unchanged. This transfer is resolved clearly in the differential absorption, \(\Delta\alpha=\alpha(T)-\alpha(15~\mathrm{K})\), which reveals a distinct feature centered at 1.66~eV whose energy remains nearly temperature independent (Fig.~\ref{fig:exciton-magnon}b), as expected for a magnon-assisted absorption sideband. This feature is also evident in the temperature dependent PLE and reflectance measurements (Supplementary Section~1 and Supplementary Fig.~2c-f). Crucially, mirroring the PL spectra (1.46~eV, 14~K) about 1.56~eV maps onto the lineshape of \(\Delta\alpha\) (1.66~eV, 300~K) (purple dashed line, Fig.~\ref{fig:exciton-magnon}c). The nearly symmetric \(\sim\)100~meV shifts of the two sidebands thus locate the underlying spin-forbidden dark exciton at 1.56~eV. As discussed in Supplementary Sections~3 and~4, this large energy scale is well beyond the calculated and measured phonon energies\cite{PRM2025Huiwen} (Supplementary Figs.~10 and~11).

Another crucial prediction of magnon-assisted sidebands is that their spectral shape and energy scale are set by the magnon and exciton dispersions. Thus, to test if the sidebands originate from the magnon-assisted process, we directly measured the magnon dispersion of \ch{La2O3Mn2Se2} by inelastic neutron scattering. Figure~\ref{fig:exciton-magnon}d displays the magnon dispersion obtained with an incident neutron energy of 40~meV at 10~K; see Supplementary Section~4 for details. A magnon plume is clearly resolved between 1.00~\AA$^{-1}$ and 2.50~\AA$^{-1}$, consistent with recent neutron scattering and resonant inelastic X-ray scattering reports\cite{asai2026, zhang2026rixs}. The one-magnon density of states (DOS), extracted from the $E_i = 60$~meV data (Supplementary Fig.~11), is shown in Fig.~\ref{fig:exciton-magnon}e (pink circles), exhibiting a broad spectral feature extending to 45~meV with a dominant peak at 31~meV and a weaker peak at 12~meV. Interestingly, the highest magnon DOS peak (\(\sim\)31~meV) lies at approximately one-third of the energy shifts of the sidebands ($\approx$ 100~meV). Thus, if these sidebands originate from magnons, they must have a substantial contribution from multi-magnon processes.

Such exchange-mediated multi-magnon processes are well established in the literature\cite{tanabe1965,sell1967,chiang1978,petrov1967,strauss1976,eremenko1974multimagnon} (Supplementary Section~5), producing sidebands that can be comparable to or stronger than their one-magnon counterparts. This behavior originates from the sudden change in the local exchange field that accompanies the recombination of the localized spin-forbidden exciton. Specifically, this couples the optical transition to a multi-spin operator rather than an isolated spin flip. To confirm this hypothesis, we extracted the exchange couplings by fitting key features of the inelastic neutron spectra to a symmetry-constrained spin model (Supplementary Section~6). The extracted parameters are then used to compute the three-magnon DOS, including only three-magnon combinations with total spin projection $\pm~\hbar$. The three-magnon DOS, shown in Fig.~\ref{fig:exciton-magnon}e (dashed curve), exhibits a broad spectral feature extending up to $\sim$130~meV, providing the characteristic energy scale for the observed sidebands. Nonetheless, to explain the detailed spectral shape, we must convolve the magnon and exciton DOS\cite{parkinson1968green,elliott1968magnon,Eremenko1974,eremenko1986}. Following standard procedures discussed in Supplementary Section~5, we have calculated this convolution and find excellent agreement with the measured spectra (orange dashed curve in Fig.~\ref{fig:exciton-magnon}c). The additional features at energies beyond our calculation likely reflect higher-order magnon processes not included here. This microscopic picture is summarized schematically in Fig.~\ref{fig:exciton-magnon}f, the characteristic sideband energy is set by the multi-magnon excitation, whereas its spectral extent reflects the combined exciton and magnon energy distributions.

\begin{figure}[]
\centering
\includegraphics[width=1\textwidth]{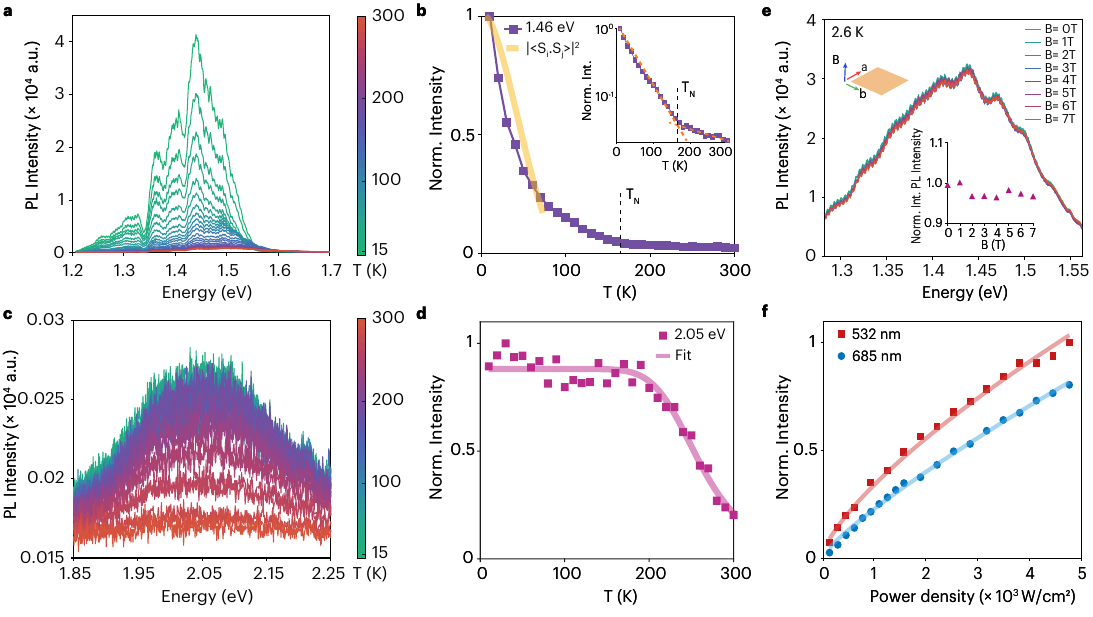}
\vspace{-4ex}
\caption{\textbf{Temperature, field and power dependence of PL.} \textbf{a,} Temperature evolution of the 1.46~eV PL band spectra measured from 10~K to 300~K. Color bar indicates the temperature. \textbf{b,} Temperature dependence of normalized integrated intensity of the 1.46~eV PL. The yellow solid line is the temperature dependence of spin-spin correlation \(|\langle S_i.S_j \rangle|^2\) calculated using spin-wave theory. The inset shows the integrated PL intensity in log-scale. The orange dashed lines are guides to the eye to show the slope change of the PL intensity in the vicinity of magnetic ordering. The vertical black dashed line indicates the N\'eel temperature, \(T_N\). \textbf{c,} Temperature evolution of the 2.05~eV PL, and \textbf{d,} corresponding integrated intensity. The solid line is the single-activation energy Arrhenius fit to the PL. \textbf{e,} Out-of-plane magnetic field (\(B_\perp\)) dependence of the low-energy PL, at 2.6~K. (Inset) integrated intensity versus B-field, indicating a spin-neutral excitation. \textbf{f,} Integrated low-energy PL intensity versus excitation power density CW 532~nm (red) and 120~fs pulsed, 685~nm (blue) at 10~K. The solid lines are the fits to the integrated PL intensity (\(I_{PL} \propto P^{\nu}\)). The absence of saturation and near-linear power dependence are inconsistent with emission from defects.} 
\label{fig:Temperature}
\end{figure}

Next we turn to the most stringent test of an exchange-mediated electric dipole process: the temperature evolution of the 1.46~eV PL intensity should closely track the underlying spin-spin correlations\cite{eremenko1986}, \(I\propto|\langle\hat{S}_i\cdot\hat{S}_j\rangle|^2\). As shown in Fig.~\ref{fig:Temperature}a, the overall PL spectral shape is nearly temperature independent, whereas its intensity depends strongly on the magnetic order. Consistent with this expectation, the integrated PL intensity (Fig.~\ref{fig:Temperature}b) rapidly decays upon heating till \(T_N\), at which point the effect of temperature is substantially reduced (inset, Fig.~\ref{fig:Temperature}b, log scale). To test if the intensity indeed scales as expected, we calculated the equal-time spin-spin correlations within spin-wave theory, including both static and dynamical contributions (described in Supplementary Section~6). To ensure the validity of this calculation, we first confirmed that the resulting temperature dependence of the static spin correlations matched well with those extracted from the measured magnetic Bragg peak intensity (Supplementary Section~4, Supplementary Fig.~12c). Returning to the PL temperature dependence, it is well reproduced by the calculated equal-time spin-spin correlations (yellow line in Fig.~\ref{fig:Temperature}b). By contrast, the PL spectra and overall intensity of the weak 2.05~eV band, shown in Fig.~\ref{fig:Temperature}c,d, remain nearly temperature independent up to \(\sim\) 200~K, well above \(T_N\). This behavior is typical of conventional PL mechanisms (band-to-band or defect-related transitions), where non-radiative pathways govern the temperature dependence. Specifically, we find the 2.05~eV band is well described by an Arrhenius behavior (Supplementary Section~1)\cite{adachi2023I,Xiaodong2018}.  

To further demonstrate that the 1.46~eV PL results from angular-momentum exchange between the spin-forbidden dark exciton and a collective magnon, we probed the PL under an applied out-of-plane magnetic field, \(B_{\perp}\). Figure~\ref{fig:Temperature}e shows the \(B_{\perp}\)-dependence of the 1.46~eV PL at 2.6~K up to 7~T (197~K data in Supplementary Fig.~14a); no measurable shift or intensity change is observed with field. This absence of a Zeeman response is expected for an exciton-magnon final state, since the sideband arises through inter-sublattice exchange rather than independent spin flips, leaving the total spin projection, \(S^{tot}_z\), unchanged and hence rendering the expected Zeeman shift negligible (discussed in detail in Supplementary Section~7)\cite{sell1966magnetic,sell1968review}. These results further support the exciton-magnon origin of the sidebands, although given the sideband's broad bandwidth and the system's strong exchange coupling, further high-field measurements are needed to confirm this.
Similarly, the PL and the 1.74~eV bright exciton reflectance band remain nearly insensitive to an in-plane field, \(B_{\parallel}\leq 7\)~T (Supplementary Fig.~14b,c), suggesting strong magnetic anisotropy. We note these responses are quite distinct from those seen in conventional magnon sidebands in non-AM systems (see Supplementary Section~8).

Finally, we addressed the possibility that the 1.46~eV PL results from defects. Specifically, one typically expects a power-law dependence on the laser excitation density: \(I_{PL}\propto P^{\nu}\), where the exponent \(\nu\) characterizes the radiative recombination, with electron-hole recombination producing \(\nu \sim 1\), and defects giving \(\nu \ll 1\) along with rapid saturation\cite{schmidt1992,jin1997}. Thus, we measured the power dependence of \(I_{PL}\) using CW (532~nm) and ultrafast (685~nm, 120~fs) lasers, as shown in Fig.~\ref{fig:Temperature}f. Notably, \(I_{PL}\) exhibits nearly linear power dependence (no saturation) with \(\nu \sim\) 0.83 (0.72) for 685~nm (532~nm) excitation laser, thereby confirming its intrinsic origin. 

To directly investigate any possible defects in \ch{La2O3Mn2Se2} crystals, we probed the local atomic arrangement of the crystals using HAADF-STEM. As shown in Extended Data~Fig. 1, the HAADF-STEM image displays an ideal crystallographic lattice with no defects or secondary phases. The clear alternating arrangement of \ch{La2O2} and \ch{Mn2OSe2} layers stacked along the [001] axis is well resolved (Extended Data Fig.~1b). The lighter oxygen columns, positioned between adjacent La columns, within the \ch{La2O2} layers are revealed in iDPC-STEM imaging (Extended Data Fig.~1c), which agrees with the expected crystal structure. Furthermore, the homogeneous spatial distribution of all elements is confirmed in the atomic-resolution STEM electron energy-loss spectroscopy (EELS) elemental mapping (Extended Data Fig.~2). The elemental maps of La, Mn, Se, and O obtained from the La \(M_{5,4}\), Mn \(L_{3,2}\), Se \(L_{3,2}\), and O \(K\) edges are shown in Extended Data Fig.~2c-f. Overall, these findings confirm the high crystalline quality of the \ch{La2O3Mn2Se2} crystals.
\section*{Discussion}

Our experimental results and first-principles calculations establish that \ch{La2O3Mn2Se2} hosts a spin-forbidden dark exciton. Crucially, its detection via the exchange-mediated electric dipole mechanism provides direct evidence for the altermagnetic state. The reciprocal-space envelope-function calculations (Supplementary Section~2) revealed that the dark exciton originates from the lowest CB and highest VB, localized near the \(X~(Y)\) high-symmetry points. Combining this with the spin character of the bands at these points, we establish the spin-forbidden nature of the dark state (Supplementary Fig.~7). Orbital-resolved band calculations (Supplementary Section~2) further reveal \ch{La2O3Mn2Se2} to be a ligand-to-metal charge-transfer insulator, with a large U \(\sim\) 2.2~eV. Because the Se states contribute only partially to the VB and not to the CB, we do not expect SOC to play a significant role in our results.

Our key experimental observation is a pair of exchange-mediated exciton-magnon sidebands from the spin-forbidden dark exciton: a Stokes-shifted PL band (1.46~eV) and an anti-Stokes-shifted absorption band (1.66~eV, Fig.~\ref{fig:exciton-magnon}c), a direct consequence of altermagnetic symmetry. This behavior contrasts sharply with dark excitons in non-magnetic systems, where spin splitting arises from relativistic SOC, and is therefore temperature independent. The detection of an opposite temperature dependence in the emission and absorption sidebands (Fig.~\ref{fig:exciton-magnon}a,b and Fig.~\ref{fig:Temperature}a,b) reflects the distinct two-particle character of the underlying exciton-magnon processes. Absorption creates an exciton and a magnon together from the crystal ground state, a process delocalized across the Brillouin zone. Its sideband strength therefore depends only on the magnetic spectral weight available at the relevant energy, set by the thermal magnon occupation. It therefore grows with temperature and persists above \(T_N\) (Fig.~\ref{fig:exciton-magnon}b, Supplementary Section~1), although some contribution from the tail of the optically allowed 1.74~eV absorption cannot be excluded. Emission, in contrast, proceeds from an already-localized exciton, which recombines via an exchange-mediated spin flip with a specific neighboring ion\cite{tanabe1965,eremenko1986}. Its intensity therefore directly tracks the equal-time spin-spin correlation (Fig.~\ref{fig:Temperature}b), consistent with spin-wave calculations of the full correlator and with an independent estimate extracted from neutron scattering (Supplementary Section~4). Consequently, its linewidth and central energy are also nearly temperature independent, in stark contrast to the emission seen in conventional magnets\cite{datta2025,kang2020coherent}. This distinguishes it further from the higher-energy 2.05~eV PL band, whose temperature dependence follows conventional mechanisms (Fig.~\ref{fig:Temperature}d).

Beyond its temperature independence, the large energy shift of the sidebands is inconsistent with a phonon-assisted mechanism. Instead, it quantitatively agrees with one- and three-magnon contributions derived from inelastic neutron scattering; their combined density of states, convolved with the exciton bandwidth, nearly reproduces the sidebands' broad spectral profile (Fig.~\ref{fig:exciton-magnon}c). The absence of a measurable Zeeman splitting in the 1.46~eV PL (Fig.~\ref{fig:Temperature}e, Supplementary Section~7) further supports its exciton-magnon sideband assignment, distinguishing this response from conventional magnon sidebands. Together, these results establish two key consequences of altermagnetic symmetry: a spin-forbidden dark exciton arising from non-relativistic spin splitting, and an exchange-mediated exciton-magnon optical transition. Both provide direct optical evidence of altermagnetic order in \ch{La2O3Mn2Se2} and, as discussed in Supplementary Section~8, are distinct from non-collinear and conventional AFMs. Beyond providing a new test of altermagnetism, these results open the door to a range of new directions. Specifically, we demonstrate a new phenomenon: optically accessing altermagnetic dark excitons via exciton-magnon transitions. We anticipate that future studies will exploit this for valley-selective excitation, coherent manipulation, and hybrid exciton-magnon quasiparticles in which excitonic and spin degrees of freedom are intrinsically entangled through exchange. More broadly, these results motivate the exploration of spin-valley physics and optically addressable spintronic and valleytronic functionality unique to insulating AMs.

\newpage
\section*
{References}
\bibliography{References}
\newpage
\newpage
\section*{Methods}
\subsection{Sample preparation}

High-quality single crystals of \ch{La2O3Mn2Se2} were prepared from the pellets of the \ch{La2O3Mn2Se2} in a high temperature tube furnace with an argon flow. The temperature was gradually increased to \(1000^\circ\)C with a ramp rate of \(3^\circ\)C/min, and then heated to \(1200^\circ\)C at a reduced rate of \(1^\circ\)C/min. The sample was annealed at \(1200^\circ\)C for 24 hours, then cooled to \(1000^\circ\)C at \(0.2^\circ\)C/min rate over 18 hours. Then cooled down to room temperature with \(3^\circ\)C/min ramp rate.

\subsection{Theoretical calculations}

 We used the open-source Quantum Espresso software package\cite{giannozzi2009} with the Perdew-Burke-Ernzerhof (PBE) generalized gradient approximation (GGA) exchange–correlation functional\cite{perdew1996} within density functional theory (DFT) to obtain the ground-state properties and wavefunctions, which were used as the starting point for the subsequent BSE calculations. The calculations were performed in a scalar-relativistic spinor formalism without spin-orbit coupling (SOC) to capture the altermagnetic ground state. We used norm-conserving pseudopotentials from PseudoDojo\cite{hamann2013}. To treat the localized Mn 3\(d\) electrons, we adopted the DFT+U scheme\cite{dudarev1998} with an effective Hubbard parameter \(U_\mathrm{eff} = 2.2\)~eV. Self-consistent and non-self-consistent DFT calculations employed a \(5\times5\times2\) Monkhorst–Pack \(k\)-grid and an 80~Ry kinetic-energy cutoff. To account for excitonic effects in the permittivity spectrum, we included electron–hole interactions by solving the Bethe-Salpeter equation (BSE) with BerkeleyGW\cite{deslippe2012,rohlfing2000}. Dielectric screening is evaluated with 1500 bands and a 20~Ry energy cutoff. For the BSE calculations, we used 12 VBs and 10 CBs and a \(10\times10\times2\) fine \(\mathbf{k}\)-grid. The resulting spectra were broadened with a 0.075 eV broadening coefficient and red-shifted by \(0.07\)~eV to compensate for the slight overestimation of the band gap compared to experiment.

\subsection{Optical measurements} The PL measurements were carried out in the backscattering configuration using a custom-built low-temperature setup\cite{tian2016low}. To excite the sample, a 532~nm (2.33~eV) CW laser was focused using a \(\text{100}\times\) objective with a spot diameter of \(\sim~\text{2}~{\mu}\text{m}\). To ensure that the PL response remains free from thermal and/or nonlinear effects, we kept the laser power low \(< \text{20}~{\mu} \text{W}\). The PL emission was dispersed by a 300 grooves per millimeter grating and collected using an Andor spectrometer coupled to an Andor charge-coupled device (CCD) detector, calibrated with Hg emission lines. The PLE measurements were performed by tuning the wavelength of the 80~MHz, 120~fs laser from 680~nm (1.82~eV) to 780~nm (1.59~eV) using the InSight ultrafast tunable laser. The laser power-dependent measurements were performed using both a CW 532~nm and a pulsed 685~nm laser, with power levels ranging from 1 to 150~\({\mu} \text{W}\), using a half-wave plate and a linear polarizer. The low-temperature magnetic-field-dependent optical measurements were performed by loading the \ch{La2O3Mn2Se2} crystal into a Quantum Design OptiCool cryostat with a base temperature of 1.6~K, applying magnetic fields up to \({\pm}7\)~T in both out-of-plane (\(B_\perp\)) and in-plane (\(B_\parallel\)) geometries relative to the crystal's \(ab\)-plane. The sample was probed through an optical window using a \(\text{100}\times\) objective lens (NA = 0.8). The PL measurements were carried out using 532~nm CW diode laser excitation (\(<\) 0.5 mW at the sample). The PL signal was collected and analyzed using a grating spectrometer (Princeton Instruments SpectraPro HRS-500) coupled to a CCD detector. Angle-resolved reflectance measurements were performed using a Fourier imaging technique by projecting the back focal plane of the \(\text{100}~\times\) objective onto the spectrometer entrance slit. The transmission and reflectance measurements were performed using an incoherent quartz–tungsten–halogen white-light source. The transmitted intensity through the sample on the sapphire substrate was normalized to the transmission measured from the bare substrate, while the reflected signal from the sample was normalized to the reflectance obtained from a silver reference mirror. The temperature dependent PL, absorption, PLE and differential reflectance measurements were done using a continuous He-flow closed-cycle cryostat (Montana Instruments).

\subsection{Neutron scattering}
Inelastic neutron scattering measurements were performed on a powder sample of \ch{La2O3Mn2Se2} using the Wide Angular-Range Chopper Spectrometer (ARCS) instrument\cite{Abernathy2012} at the Spallation Neutron Source of Oak Ridge National Laboratory. The sample (mass 2.416~g) was loaded into a cylindrical aluminum sample can, and inelastic scattering spectra were collected at 10~K using an incident neutron energy of 40~meV and 60~meV with instrumental settings optimized for energy resolution\cite{LIN201926}. Additionally, the instrument's event data were used in a continuous cooling run (T = 238~K to 10~K) and subsequently binned into 5~K intervals\cite{Granroth2018}. For the neutron measurements, the sample was enclosed in an Al can, where the instrumental background was measured using the empty portion of the sample can and subtracted from the data to isolate the signal from the sample. The neutron data reduction was performed with the Mantid package\cite{mantid2014}. The ARCS instrument has a large array of pixelated detectors. To remove any variation in detector efficiency and to correct for different solid angle coverage, the neutron signal is normalized to a white beam vanadium measurement that was taken at the beginning of the run cycle\cite{osti_3014817}.

\subsection{Electron microscopy}

Cross-section STEM specimens were prepared using a Thermo Fisher Scientific Helios G4 UX focused ion beam (FIB) using standard lift-out and thinning methods. Electron microscopy data were acquired using a Cs-corrected Thermo Fisher Scientific Spectra 300 X-CFEG operating at 300~kV. A convergence angle was 30~mrad, and a collection angle range of 60-200~mrad for HAADF-STEM imaging, and 6-29~mrad for iDPC-STEM using a four-quadrant detector. To enhance the signal-to-noise ratio, multiple fast-acquisition images were acquired, then aligned and summed using a rigid registration process optimized for noisy images.

\subsection{Data availability}
All the data supporting the findings of this study are available.

\subsection{Acknowledgments}
The work of K.S.B. (project supervision), B.S. and K.-M.K. (optical constants and low-temperature PL) was supported by the U.S. Department of Energy (DOE), Office of Science, Basic Energy Sciences (BES) under Award \#SC0018675. The measurements of PLE by Y.Z. were supported by the Air Force Office of Scientific Research under award number FA9550-24-1-0110. First principles calculations by X.X. and D.Y.Q. were supported by National Science Foundation Division of Chemistry under Award No. CHE-2412412. Development of the BerkeleyGW code was supported by the Center for Computational Study of Excited-State Phenomena in Energy Materials (C2SEPEM) at the Lawrence Berkeley National Laboratory, funded by the U.S. Department of Energy, Office of Science, Basic Energy Sciences, Materials Sciences and Engineering Division, under Contract No. DE-AC02-05CH11231. The calculations used resources of the National Energy Research Scientific Computing (NERSC), a DOE Office of Science User Facility operated under Contract No. DE-AC02-05CH11231, under Award No. BES-ERCAP-0031507 and No. BES-ERCAP-0027380. An award of computer time was provided by the U.S. Department of Energy’s (DOE) Innovative and Novel Computational Impact on Theory and Experiment (INCITE) Program. This research used resources from the Argonne Leadership Computing Facility, a U.S. DOE Office of Science user facility at Argonne National Laboratory, which is supported by the Office of Science of the U.S. DOE under Contract No. DE-AC02-06CH11357. This material is based upon research conducted by D.Y.Q. as an Alfred P. Sloan Research Fellow under Grant No. FG-2026-79515. B.A.F. and S.R.H. (inelastic neutron scattering) were supported by DOE-BES through Award No. DE-SC0021134. The National Science Foundation, Award No. DMR-2310895, supported the work of M.G. (device fabrication). A portion of this research used resources at the Spallation Neutron Source, a DOE Office of Science User Facility operated by the Oak Ridge National Laboratory. The beam time was allocated to ARCS on proposal number IPTS-36223. STEM characterization was supported by the DOE BES DE-SC0023905, using electron microscopes supported by the Platform for the Accelerated Realization, Analysis, and Discovery of Interface Materials (PARADIM) NSF Award No. DMR-2039380, and the Cornell Center for Materials Research shared instrumentation facility. The contribution of C.-C.W. and H.J. was supported by an NSF Career grant 2145832. X.L. and F.L. acknowledge financial support from the DOE-BES (No. DE-FG02-04ER46148) and computational resources from CHPC of the University of Utah and the DOE-NERSC.

\subsection{Author contributions} B.S. performed the optical measurements and analyzed the data. K.-M.K., Y.Z., M.S., C.G., A.L. and Q.M. helped with optical data collection. X.X. and D.Y.Q. performed the DFT calculations. V.W. and B.F. performed the theoretical modeling of the magnon DOS. S.R.H., G.E.G., and B.A.F. performed the inelastic neutron scattering measurements. S.P. and V.M.M. performed the magnetic-field dependent PL measurements. M.G. helped with sample preparation for the optical measurements. X.L. and F.L. calculated the phonon dispersion. Y.-M.W. and J.J.C. performed the STEM measurements. C.-C.W. and H.J. synthesized the \ch{La2O3Mn2Se2} single crystals. B.S. wrote the paper with the help of K.S.B. K.S.B. conceived and supervised the project. All authors contributed to the discussions about the paper.    

\subsection{Competing interests}
The authors declare no competing interests.
\clearpage

\begin{figure}[]
\centering
\includegraphics[width=1\textwidth]{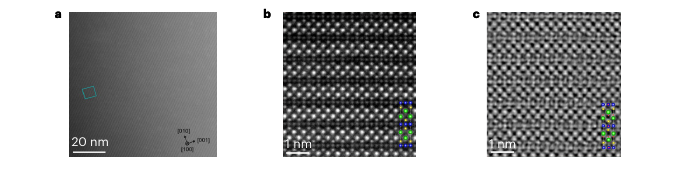}
\caption*{\textbf{Extended Data Fig. 1:} \textbf{HAADF- and iDPC-STEM images of \ch{La2O3Mn2Se2}.} \textbf{a,} Single crystal viewed along the [100] direction, showing no observable defects or secondary phases. Scale bar 20~nm. \textbf{b,c,} High-magnification HAADF-STEM \textbf{(b)}, and iDPC-STEM (integrated differential phase contrast) \textbf{(c)} images of the region marked by the cyan rectangle in \textbf{(a)}, resolving the atomic arrangement of alternating \ch{La2O2} and \ch{Mn2OSe2} layers. Scale bars, 1~nm.}
\end{figure}

\begin{figure}[]
\centering
\includegraphics[width=1\textwidth]{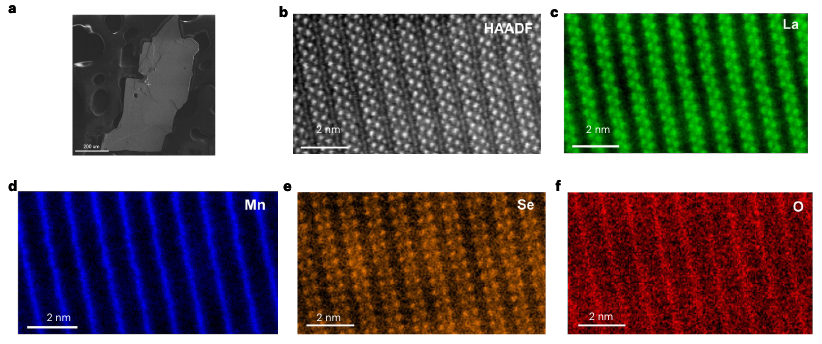}
\caption*{\textbf{Extended Data Fig. 2: Atomic-resolution STEM-EELS elemental maps of \ch{La2O3Mn2Se2} single crystal.} \textbf{a,} 
SEM-secondary electron (SE) image of \ch{La2O3Mn2Se2} crystal. \textbf{b,} HAADF-STEM image acquired along the [100] zone axis. \textbf{c-f,} The corresponding elemental maps of La (c), Mn (d), Se (e), and O (f), obtained from the La \(M_{5,4}\), Mn \(L_{3,2}\), Se \(L_{3,2}\), and O \(K\) edges, respectively.}
\end{figure}
\end{document}